\documentclass[a4paper,twoside,reqno]{bjp}
\usepackage{graphicx}
\usepackage{cite}
\usepackage{amssymb,amsmath,amscd,amsthm}
\usepackage{times}

\usepackage[bookmarks=false]{hyperref}
\hypersetup{%
    colorlinks=true,        
    linkcolor=blue,          
    citecolor=blue,         
    urlcolor=blue           
    }

\usepackage{geometry}
\allowdisplaybreaks

\begin{document}

\title{Microscopic version of the Bohr-Mottelson model and its application}

\runningheads{H. G. Ganev}{Microscopic version of the Bohr-Mottelson
model and its application}

\begin{start}{%
\author{H. G. Ganev}{1},

\address{Joint Institute for Nuclear Research, Dubna, Russia}{1}

\received{Day Month Year (Insert date of submission)}
}

\begin{Abstract}
The shell-model coupling scheme of the proton-neutron symplectic
model (PNSM), defined by the following dynamical symmetry chain
$Sp(12,R) \supset SU(1,1) \otimes SO(6) \supset U(1) \otimes
SU_{pn}(3) \otimes SO(2) \supset SO(3)$, is considered. It is shown
that it corresponds to a microscopic version of the Bohr-Mottelson
collective model which captures the original relationships between
its exactly solvable submodel limits. This variant of the PNSM
provides an interesting and relevant shell-model symplectic-based
framework for exploring the nuclear collective dynamics. Some simple
applications of the present theory to different nuclei with various
collective properties are given.
\end{Abstract}

\begin{KEY}
Bohr-Mottelson model, proton-neutron symplectic model, Sp(12,$R$)
dynamical algebra
\end{KEY}
\end{start}


\section{Introduction}

It is well known that in nuclear physics there are two fundamental
models of nuclear structure that have been awarded with Nobel
Prizes. The first one is the Bohr-Mottelson (BM) collective model
\cite{BM} which is based on the quantization of the classical
picture of surface vibrations and rotations. It has influenced all
collective models because it provides the basic concepts and
language in terms of which the nuclear collective motion can be
described. It has demonstrated that the low-lying nuclear states can
be described by considering only few macroscopic collective degrees
of freedom when the intrinsic excitations lie high in energy. The
second model is the shell model \cite{Mayer49,Heyde90} which
includes all many-fermion degrees of freedom. It provides a general
microscopic framework in terms of which the other collective models
can be founded and expressed.

The problem of giving the BM model a microscopic foundation have
been realized long time ago. Its solution, however, was given
through the algebraic approach by embedding it in the nuclear shell
model. It was shown (see, e.g. \cite{stretched,Rowe96}) that the
collective model of Bohr and Mottelson admits a microscopic
realization first by augmenting it by vorticity degrees of freedom,
important for the appearance of low-lying collective states, and
second by making it compatible with the composite many-fermion
structure of the nucleus. The result is the one-component $Sp(6,R)$
symplectic model \cite{RR1} of nuclear collective motion, sometimes
called a microscopic collective model, which is a submodel of the
nuclear shell model. The $Sp(6,R)$ model of nuclear rotations, among
its submodels, contains the rigid-rotor model \cite{rot3} and the
Elliott's $SU(3)$ shell model of collective rotations
\cite{Elliott58}, which obviously can be associated only with the
rotor model limit of the BM model. The presence of vorticity in the
$Sp(6,R)$ model results in a complete range of possible collective
flows from irrotational-flow (zero vorticity) to rigid rotations.
This is of significant importance as well as the fact that the
vortex-spin degrees of freedom are responsible for the appearance of
low-lying collective states \cite{stretched,Rowe96}. However, the
$Sp(6,R)$ model does not contain an $SO(5)$ or $SO(6)$ structure,
which could allow to associate it with the $\beta$-rigid or
$\beta$-soft but $\gamma$-unstable type dynamics of the Wilets-Jean
(WJ) \cite{WJ} model, in a manner similar, e.g., to that of IBM
\cite{IBM}.

Recently, the generalized Bohr-Mottelson model was embedded in the
two-component proton-neutron microscopic shell-model theory of the
nucleus \cite{microBM} within the framework of the proton-neutron
symplectic model (PNSM) \cite{cdf,smpnsm}, providing in this way a
more natural interpretation of the underlying BM quadrupole-monopole
collective dynamics. The new embedding is performed by considering
the shell-model coupling scheme of the PNSM, defined by the
following dynamical symmetry chain $Sp(12,R) \supset SU(1,1) \otimes
SO(6) \supset U(1) \otimes SU_{pn}(3) \otimes SO(2) \supset SO(3)$
\cite{microBM}. We will demonstrate here that, in contrast to the
$Sp(6,R)$ symplectic model and the popular IBM \cite{IBM}, the
presently considered microscopic version of the BM model has exactly
solvable limits that have a relationship, closely resembling the one
between the original $\gamma$-unstable WJ \cite{WJ} and rigid-rotor
\cite{DF,rot3} submodels.

\section{Bohr-Mottelson model and its exactly solvable limits}

The BM model can be formulated in algebraic terms by means of
different spectrum generating algebras (SGA) and dynamical groups,
which allows easily to perform an analysis at the algebraic level
and to establish a relationship with other collective models of
interest. The building blocks of the BM model are provided by the
position $\alpha_{\mu}$ and momentum $\pi^{\nu}$ coordinates, which
together with the identity operator $I$ close the Lie algebra of
Heisenberg-Weyl group $HW(5)=\{\alpha_{\mu},\pi^{\nu},I\}$. In terms
of quadrupole phonon operators, one can use the alternative
realization $HW(5)=\{d^{\dag}_{\mu}, d_{\nu}, I\}$. Different
dynamical groups and SGA can be constructed from these coordinates.
For our purposes, the following two dynamical groups $[HW(5)]U(5)$
and $SU(1,1) \otimes SO(5)$ \cite{RW} are of interest. It has been
shown \cite{RW,Rowe-ACM1} that the three dynamical subgroups $U(5)$,
$[R^{5}]SO(5)$ and $[R^{5}]SO(3)$ correspond to the spherical
vibrator, $\gamma$-unstable Wilets-Jean \cite{WJ} and rigid-rotor
\cite{DF,rot3} exactly solvable limits of the BM model,
respectively.

\subsection{Harmonic spherical vibrator}

The states in this limit are classified by the states of the
five-dimensional harmonic oscillator, defined by the reduction
chain:
\begin{equation}
[HW(5)]U(5) \supset U(5) \supset SO(5) \supset SO(3), \label{HV-DS}
\end{equation}
where $[HW(5)]U(5)$ is the semi-direct product group of
Heisenberg-Weyl group, $HW(5)$ and $U(5)
=\{d^{\dag}_{\mu}d_{\nu}\}$, the latter being the symmetry group of
the oscillator. Then for the HO Hamiltonian $H_{HV} =
\sum_{\mu}\big(d^{\dag}_{\mu}d_{\mu} +\frac{5}{2}\big)$ one
immediately obtains the energies given by $E_{N}=\big(N+5/2\big)
\hbar\omega$. The components of $E2$ transition operator
$T^{E2}\simeq \alpha_{\mu}$ are provided by the generators of
Abelian subgroup $R^{5} \equiv
\{\alpha_{\mu}=\frac{1}{\sqrt{2}}(d^{\dag}_{\mu}+d_{\mu});
[\alpha_{\mu},\alpha_{\nu}] = 0 \}$ of $HW(5)$.

\subsection{Wilets-Jean model}

The dynamical subgroup chain of the $\gamma$-unstable $\beta$-rigid
WJ model is \cite{Rowe-ACM1}:
\begin{equation}
[HW(5)]U(5) \supset [R^{5}]SO(5) \supset SO(5) \supset SO(3),
\label{WJ-DS}
\end{equation}
where $[R^{5}]SO(5)$ is the semi-direct product group of $R^{5}$ and
$SO(5)=\{\Lambda_{\mu\nu}=-i(d^{\dag}_{\mu}d_{\nu}-d^{\dag}_{\nu}d_{\mu})\}$.
The irreducible representations of $[R^{5}]SO(5)$ are characterized
by the rigid values of $\beta =\beta_{0}$. Hence, there is a problem
with the delta-function nature of the $\beta$ wave functions, which
in turn don't have a convergent expansion in terms of the harmonic
oscillator $U(5)$ states.

The $E2$ transition operator $T^{E2}\simeq \alpha_{\mu}$ in the WJ
model belongs to $R^{5}$, while the Hamiltonian is expressed as a
linear combination of the second-order Casimir operator of $SO(5)$,
i.e. $H_{WJ} = A'C_{2}[SO(5)]=A'\Lambda^{2}$. It eigenvalues are
then given in terms of the $SO(5)$ quantum number $\tau$ by
$E_{\tau}=A'\tau(\tau+3)$. The yrast levels having $L = 2\tau$
produce a characteristic ratio $E_{4^{+}_{1}}/E_{2^{+}_{1}} = 2.50$
of the WJ $\gamma$-unstable model for the first $2^{+}$ and $4^{+}$
states.

\subsection{Rigid rotor model}

The dynamical subgroup chain of the $\beta$-rigid and $\gamma$-rigid
rotor model is \cite{Rowe-ACM1}:
\begin{equation}
[HW(5)]U(5) \supset [R^{5}]SO(5) \supset [R^{5}]SO(3) \supset SO(3),
\label{RR-DS}
\end{equation}
where $ROT(3) \equiv [R^{5}]SO(3) = \{L_{k},
\alpha_{\mu}|[\alpha_{\mu},\alpha_{\nu}]=0 \}$ is the rigid-rotor
model group of Ui \cite{rot3}. The irreducible representations of
the $ROT(3)$ group are characterized by both $\beta$-rigid and
$\gamma$-rigid values. Looking at Eqs.(\ref{WJ-DS}) and
(\ref{RR-DS}), it follows that the $\beta$-rigid $\gamma$-rigid
rotor model is a submodel of the WJ $\beta$-rigid but
$\gamma$-unstable model since the $[R^{5}]SO(3)$ is a subgroup of
$[R^{5}]SO(5)$. This relationship between the two submodels has not
been widely exploited in the literature. Hence, in the BM
rigid-rotor submodel again there is a problem with the wave
functions which are delta functions in both $\beta$ and $\gamma$.

Exact solution exists for the case of an axially-symmetric rotor.
The rigid rotor Hamiltonian $H_{rot} =
\sum^{3}_{k=1}\frac{\hbar^{2}\overline{L}^{2}_{k}}{2J_{k}}$ for this
case becomes $H_{rot} =
\frac{\hbar^{2}L^{2}}{2J_{1}}+\big(\frac{\hbar^{2}}{2J_{3}}
-\frac{\hbar^{2}}{2J_{1}}\big)\overline{L}^{2}_{3}$, which
eigenvalues are given by $E_{KL} =
\frac{\hbar^{2}}{2J_{1}}L(L+1)+\big(\frac{\hbar^{2}}{2J_{3}}
-\frac{\hbar^{2}}{2J_{1}}\big)K^{2}$. The set $\{\overline{L}_{k}\}$
labels the intrinsic $SO(3)$ angular momentum operators, $J_{k} =
4\mathfrak{B}\beta^{2}_{0}$ $sin^{2}(\gamma_{0}-2\pi k/3)$ are the
irrotational-flow moments of inertia, and $K$ is the third
projection of the angular momentum operator on the body-fixed axis
$3$. Usually, in the numerical applications, the moments of inertia
$J_{k}$ are treated as free parameters that are fitted to the
experimental data. In this way, we see that the Hamiltonian of an
axially-symmetric rotor is expressed in terms of the Casimir
operators in the chain $SO(3) \supset SO(2)$. The $E2$ transition
operator $T^{E2}\simeq \alpha_{\mu}$ is an element of the
rigid-rotor algebra since $[R^{5}]SO(3) = \{L_{k}, \alpha_{\mu}\}$.
Its matrix elements are proportional to the ordinary $SO(3)$
Clebsch-Gordan coefficients.

\subsection{Algebraic collective model}

Although the last two BM limiting cases just considered are
characterized by dynamical subgroup chains, they are not
particularly useful for the construction of basis states in which to
diagonalize more general collective Hamiltonians, as this is done in
the case of the five-dimensional oscillator. This is because the
wave functions which diagonalize the $[R^{5}]SO(5)$ and
$[R^{5}]SO(3)$ subgroups are not square-integrable. They contain
factors which are delta functions in $\beta$ and $\gamma$. This
limitation expresses the fact that rigidly-defined intrinsic
quadrupole moments are unphysical and incompatible with the quantum
mechanics. The resolution of this problem in the ACM is obtained by
relaxing the $\beta$-rigidity of WJ model by replacing its dynamical
group $[R^{5}]SO(5)$ with $SU(1,1) \otimes SO(5)$, which results in
a more physical collective model. Thus, in the ACM the following
dynamical symmetry chain is used to define a continuous set of basis
states for the BM model \cite{Rowe-ACM1,Rowe-ACM2}:
\begin{align}
&SU(1,1) \otimes SO(5) \supset U(1) \otimes SO(3) \supset SO(2),
\label{ACM-DS} \\
&\qquad \lambda_{\upsilon} \qquad\quad \upsilon \quad \ \alpha \quad
n \qquad\quad L \qquad\quad M \notag
\end{align}
where $SU(1,1)$ is a dynamical group for radial $\beta$ wave
functions and $SO(5)$ group determines the angular part ($SO(5)$
spherical harmonics) that is characterized by the seniority quantum
number $\upsilon$. An important characteristic of the ACM is that it
enables $\beta$-rigid and $\gamma$-rigid limits to be approached in
a continuous way with increasingly narrow but nevertheless
square-integrable $\beta$ and $\gamma$ wave functions. The WJ and
rigid-rotor submodels of the BM model are then seen as special cases
of the more physical ACM. Additionally, the energies in the WJ and
harmonic vibrator limits can be written as
\begin{equation}
E_{WJ}=A\upsilon(\upsilon+3), \label{WJenergies}
\end{equation}
and
\begin{equation}
E(n,\upsilon)=\bigg(2n+\upsilon+\frac{5}{2}\bigg)\hbar\omega,
\label{HVenergies}
\end{equation}
respectively.

\section{PNSM shell-model classification of the collective states}

In the present work, we classify the shell-model nuclear states
within the PNSM by the following reduction chain \cite{microBM}:
\begin{align}
&Sp(12,R) \supset SU(1,1) \otimes SO(6) \supset U(1) \otimes
SU_{pn}(3) \otimes SO(2) \supset SO(3), \label{Sp2RxO6-DS} \\
&\qquad\langle\sigma\rangle \qquad\qquad \lambda_{\upsilon}
\qquad\quad \upsilon \qquad\quad \ p \qquad \ (\lambda,\mu)
\qquad\quad \nu \quad \ \ q \quad \ L \notag
\end{align}
where bellow the different subgroups are given the quantum numbers
that characterize their irreducible representations. The basis
functions along the chain (\ref{Sp2RxO6-DS}) can thus be written in
the form \cite{microBM}:
\begin{equation}
\Psi_{\lambda_{\upsilon}p;\upsilon\nu qLM}(r,\Omega_{5}) =
R^{\lambda_{\upsilon}}_{p}(r)Y^{\upsilon}_{\nu qLM}(\Omega_{5}),
\label{SU11xSO6wf}
\end{equation}
where $Y^{\upsilon}_{\nu qLM}(\Omega_{5})$ are the SO(6) Dragt's
spherical harmonics \cite{Dragt65,Chacon84}. The $SU(1,1)$ is a
dynamical group for radial wave functions and $SO(6)$ group
determines the angular part ($SO(6)$ spherical harmonics) that is
characterized by the seniority quantum number $\upsilon$.

The $SU(1,1)$ algebra has unitary representations with orthonormal
basis states $\{|\lambda_{\upsilon}, p \rangle; p = 0, 1, 2,
\ldots\}$ that are defined by the equations \cite{sp2rxso6}:
\begin{align}
&S^{(\lambda_{\upsilon})}_{+} |\lambda_{\upsilon}, p \rangle =
\sqrt{(2\lambda_{\upsilon}+p)(p+1)}
|\lambda_{\upsilon}, p+1 \rangle, \label{AS+}\\
&S^{(\lambda_{\upsilon})}_{-} |\lambda_{\upsilon}, p \rangle =
\sqrt{(\lambda_{\upsilon}-1+p)p} |\lambda_{\upsilon}, p-1 \rangle,
\label{AS-}\\
&S^{(\lambda_{\upsilon})}_{0} |\lambda_{\upsilon}, p \rangle =
\frac{1}{2}(\lambda_{\upsilon} + 2p) |\lambda_{\upsilon}, p \rangle,
\label{AS0}
\end{align}
for any value of $\lambda_{\upsilon}$ ($\lambda_{\upsilon} > 1$).
Thus, e.g., for the harmonic oscillator series representations of
$SU(1,1)$ we have $\lambda_{\upsilon} = \upsilon + 6/2$. From
Eq.(\ref{AS0}) it then follows $E_{p\upsilon} = (2p +
\lambda_{\upsilon})\hbar\omega = (2p + \upsilon + 6/2)\hbar\omega$
to be compared with Eq.(\ref{HVenergies}). For $\lambda_{\upsilon} >
\upsilon + 6/2$, the radial wave functions are eigenfunctions of a
Hamiltonian with a potential \cite{Rowe-Euclidean,Rowe06c}
\begin{equation}
V^{(\lambda_{\upsilon})}(r) =
\frac{(\lambda_{\upsilon}-1)^{2}-4}{2r^{2}}+\frac{r^{2}}{2},
\label{DP}
\end{equation}
which corresponds to the addition of a centrifugal-like potential to
the harmonic oscillator potential. In this case,
$\lambda_{\upsilon}=1+\sqrt{(\upsilon+4/2)^{2}+(r_{0})^{4}}$ and one
obtains the so-called modified oscillator $SU(1,1)$ irreps
\cite{Rowe-Euclidean}. The energies for the Davidson oscillator
$E_{p\upsilon} = (2p+\lambda_{\upsilon})\hbar\omega =
\Big(2p+1+\sqrt{(\upsilon+4/2)^{2}+(r_{0})^{4}}\Big)\hbar\omega$ for
large values of $r_{0}$ can then be expanded in inverse powers of
$r_{0}$ to give
\begin{equation}
E_{p\upsilon} = E_{0} + 2p \hbar\omega + A\upsilon(\upsilon+4) +
\ldots \label{EnExpan}
\end{equation}
where $A=\hbar\omega/2r_{0}^{2}$. For $r_{0} \rightarrow \infty$ and
$\hbar\omega \rightarrow \infty$ this expression corresponds to the
microscopic counterpart of the $\beta$-rigid, $\gamma$-unstable WJ
model with energies given by $E_{\upsilon} = E_{0} +
A\upsilon(\upsilon+4)$. Note that, in contrast to WJ model value
2.50, this expression gives a characteristic ratio
$E_{4^{+}_{1}}/E_{2^{+}_{1}} \simeq 2.67$ of the ground state band
energies, for which $L=\upsilon$ (see the left diagonal of Table 1
of Ref. \cite{microBM} with $(\lambda,\mu)=(k,0)$, $k=0, 2, 4,
\ldots$).

A microscopic analogue of the rigid rotor is provided by the
$SU_{pn}(3)$ structure, which for large dimensional $SU(3)$
representations contracts \cite{su3rot3} to $[R^{5}]SO(3)$ of Ui and
the rigid rotor model states are approached.

The results just obtained can alternatively be achieved in a pure
algebraic way by taking the proper Casimir operators in the
Hamiltonian, as it is done in the next section.

\section{Application}

Applications within the framework of the algebraic approach to
nuclear structure vary according to the type of nuclear interaction
that is used. Initially, the simplest case of using schematic or
algebraic interactions has been widely exploited by many authors.
But with the increase of the high-performance computing technologies
during the last decades, different computationally intensive
large-scale calculations that use modern high-precision realistic
interactions inspired from the QCD became possible, like those
performed within the no-core shell model (NCSM) \cite{NCSM} or
symmetry-adapted no-core shell model (SA-NCSM) \cite{SA-NCSM}
frameworks. Here, however, we use a simple algebraic interaction.

A general dynamical symmetry Hamiltonian can be written as a linear
combination of the Casimir operators of different subgroups of the
chain (\ref{Sp2RxO6-DS}):
\begin{align}
H = \
&H\Big(S^{(\lambda_{\upsilon})}_{0},S^{(\lambda_{\upsilon})}_{+},S^{(\lambda_{\upsilon})}_{-}\Big)
+ V(r) \notag\\
&+ A\Lambda^{2} + f\Big(C_{2}[SU_{pn}(3)],C_{2}[SO(3)]\Big),
\label{rsoft-H}
\end{align}
where
$\{S^{(\lambda_{\upsilon})}_{0},S^{(\lambda_{\upsilon})}_{+},S^{(\lambda_{\upsilon})}_{-}\}$
are the generators of the group $SU(1,1)$ and
$\Lambda^{2}=C_{2}[SO(6)]$. The starting point of the present
application is the following dynamical symmetry Hamiltonian
\begin{align}
H = &n \hbar\omega + A\Lambda^{2} + BC_{2}[SU_{pn}(3)]  \notag\\
&+ aC_{2}[SO(3)] + bK^{2} + c(C_{2}[SO(3)])^{2}, \label{SU11xSO6-H}
\end{align}
where we have used the fact that $H_{0} =2S^{(\lambda)}_{0} = n
\hbar\omega$ represents the harmonic oscillator mean field. The last
three terms represent a residual rotor part which takes into account
the band characteristics, like the observed moment of inertia, the
$K$-band splitting and the centrifugal stretching effects.
Additionally, we consider the following simple Hamiltonian
\cite{sp2rxso6}:
\begin{align}
&H_{hmix} = h\Big(G^{2}(a,a) \cdot F^{2}(b,b) + G^{2}(b,b) \cdot
F^{2}(a,a)\Big),  \label{Hmix}
\end{align}
which mixes different $SU_{pn}(3)$ multiplets within the maximal
seniority $SO(6)$ representation $\upsilon_{0}$ contained in the
corresponding symplectic bandhead. For more information we refer the
reader to Ref.\cite{sp2rxso6}, where the required matrix elements
are also given.

\begin{figure}[h!]\centering
\includegraphics[width=60mm]{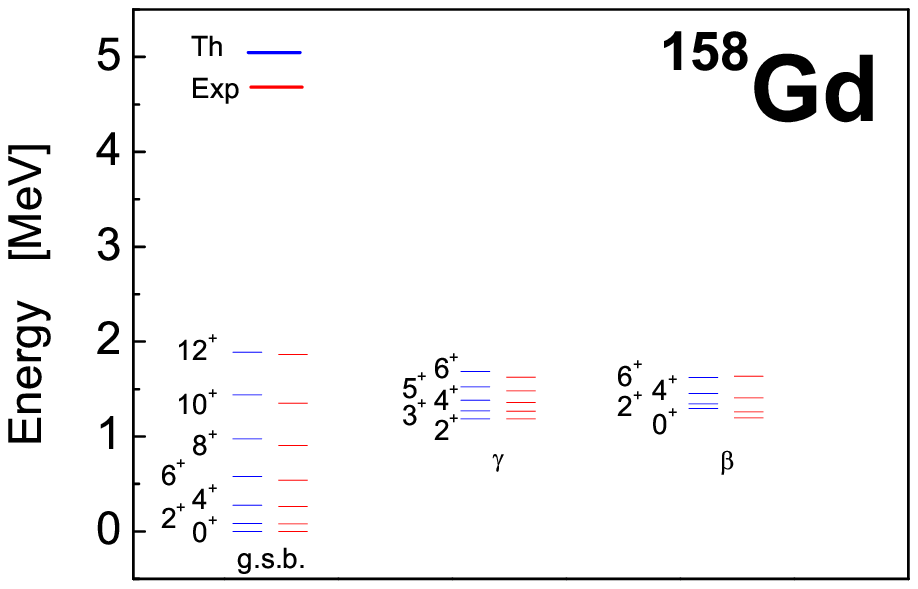}\vspace{5.mm}
\includegraphics[width=60mm]{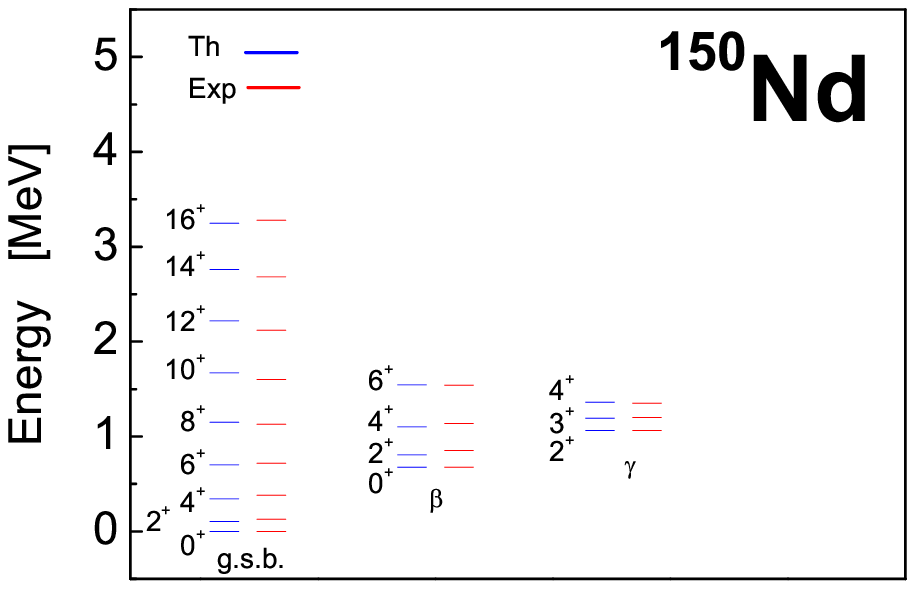}\vspace{5.mm}
\includegraphics[width=60mm]{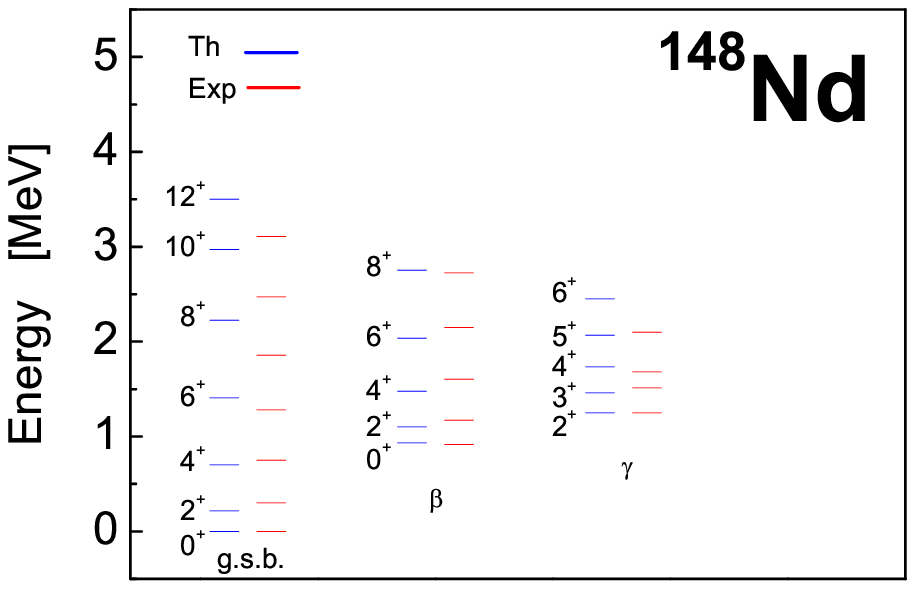}
\caption{Comparison of experimental energy levels with the theory
for the low-lying ground, $\beta$ and $\gamma$ bands in $^{158}$Gd,
$^{150}$Nd, and $^{148}$Nd.}\label{spectra}
\end{figure}

\begin{figure}[h!]\centering
\includegraphics[width=60mm]{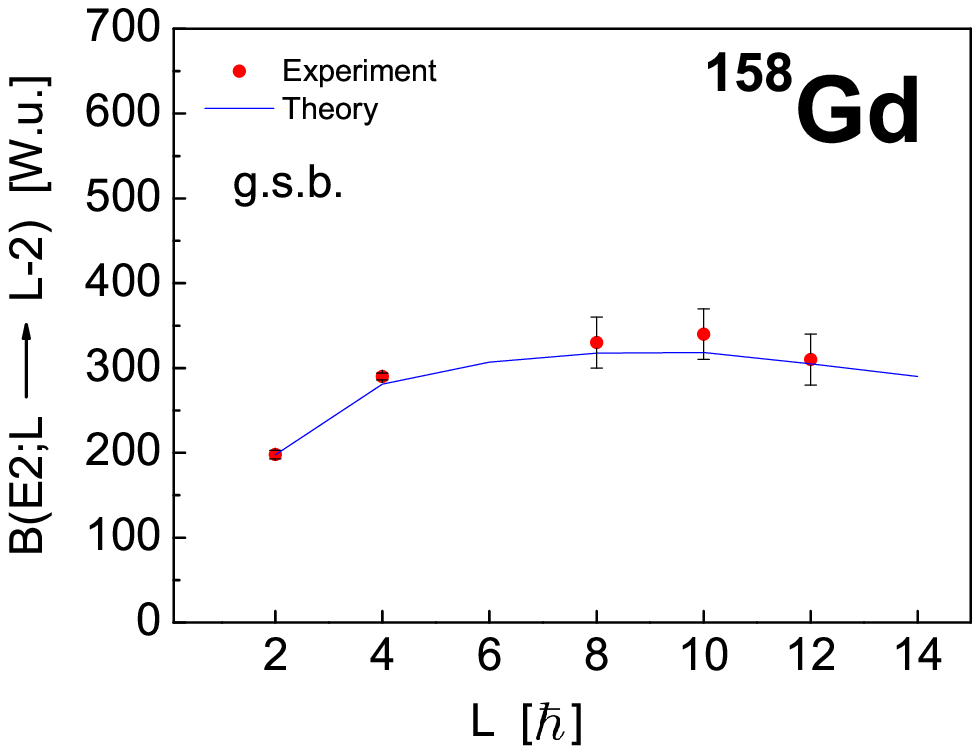}\vspace{1.mm}
\includegraphics[width=60mm]{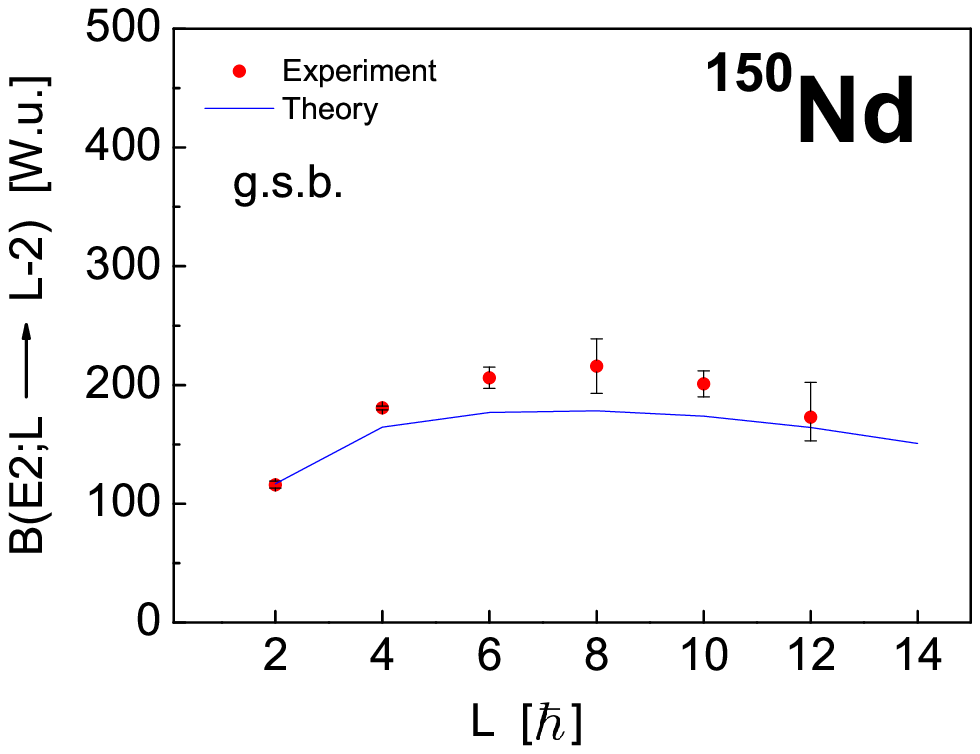}\vspace{1.mm}
\includegraphics[width=60mm]{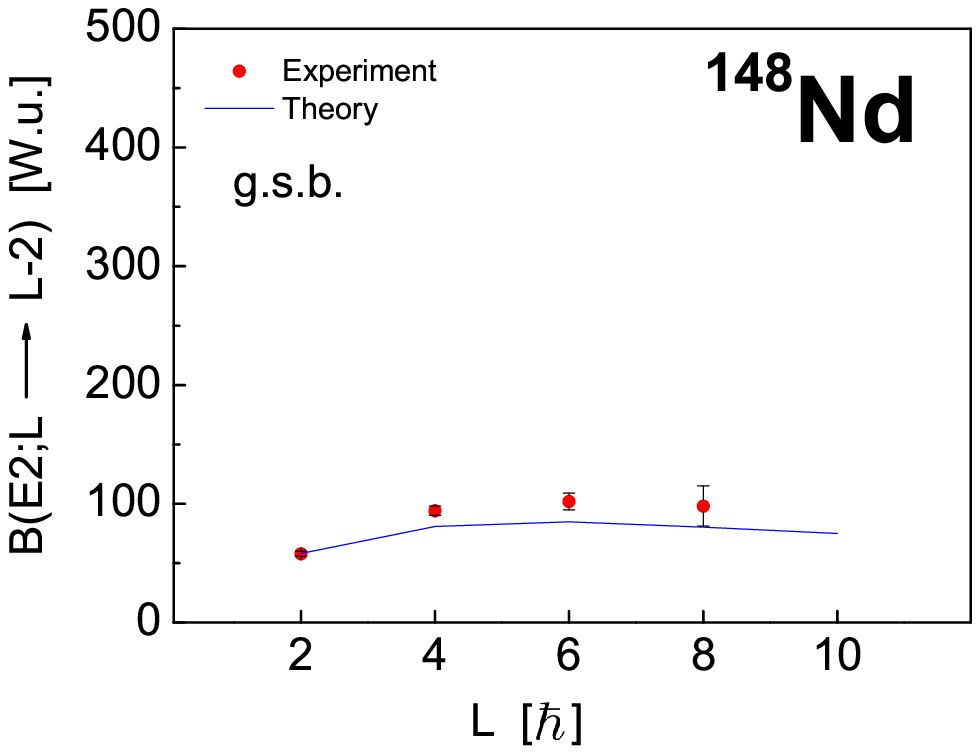}
\caption{Calculated intraband $B(E2)$ values in Weisskopf units
between the states of the ground band in $^{158}$Gd, $^{150}$Nd, and
$^{148}$Nd. No effective charge is used.} \label{be2s}
\end{figure}

We consider the following three nuclei $^{158}$Gd, $^{150}$Nd and
$^{148}$Nd with the characteristic ratios \cite{exp}
$E_{4^{+}_{1}}/E_{2^{+}_{1}} \simeq 3.26$, $2.93$, and $2.49$,
respectively. The relevant $Sp(12,R)$ irreducible representation for
each nucleus is determined by the lowest-weight $U(6)$ irrep, which
in turn is fixed by the underlying proton-neutron shell-model
structure. Thus, according to the shell-model considerations based
on the pseudo-$SU(3)$ scheme \cite{pseudoSU3c}, we choose the
following $Sp(12,R)$ irreducible representations: 0p-0h $[36]_{6}$
for $^{158}$Gd, 0p-0h $[24]_{6}$ for $^{150}$Nd, and 0p-0h
$[18]_{6}$ for $^{148}$Nd, respectively. We diagonalize the model
Hamiltonian in the irreducible collective space spanned by the the
maximal seniority $SO(6)$ representation $\upsilon_{0}$, i.e.
$\upsilon_{0}=36$ ($^{158}$Gd), $\upsilon_{0}=24$ ($^{150}$Nd), and
$\upsilon_{0}=18$ ($^{148}$Nd), correspondingly, of the symplectic
bandhead for each nucleus. Hence the first two terms in the
Hamiltonian (\ref{SU11xSO6-H}) are irrelevant and can be dropped. In
addition, due to the prolate-oblate symmetry of the $SU_{pn}(3)$
multiplets related with the conjugate SU$_{pn}$(3) multiplets
$(\lambda,\mu)$ and $(\mu,\lambda)$ contained within the
corresponding SO(6) irreducible representations, we use only the
SU(3) multiplets $(\lambda,\mu)$ with $\lambda \geq \mu$. The
results of diagonalization for the low-lying excitation spectra in
$^{158}$Gd, $^{150}$Nd, and $^{148}$Nd together with the
experimental data are shown in Fig. \ref{spectra}, while the
intraband $B(E2)$ transition strengths between the states of the
ground band for these three nuclei are given in Fig. \ref{be2s}. In
the calculation of the corresponding $B(E2)$ values, no effective
charges are used. The values of the model parameters (in MeV) are as
follows: $B = -0.039$, $a = 0$, $b = 0.247$, $c = 0$, and $h =
-0.147$ for $^{158}$Gd; $B = -0.046$, $a = 0.015$, $b = 0.075$, $c =
-0.000016$ and $h = -0.074$ for $^{150}$Nd; $B = -0.038$, $a =
0.023$, $b = 0.203$, $c = -0.000055$ and $h = -0.0924$ for
$^{148}$Nd. We see a good description of the experimental data for
all the three nuclei under consideration. Note also that the
excitation spectra of $^{158}$Gd is obtained without using an
adjustable moment of inertia.

\section{Conclusions}

Microscopic analogues of the exactly solvable limits of the
Bohr-Mottelson collective model are shortly considered in respect to
the original BM submodels. Microscopic considerations are given
within the framework of the PNSM, in which the relevant shell-model
coupling scheme is defined by the following dynamical symmetry chain
$Sp(12,R) \supset SU(1,1) \otimes SO(6) \supset U(1) \otimes
SU_{pn}(3) \otimes SO(2) \supset SO(3)$, according to which the
many-particle nuclear shell-model states are classified. We note
that within the present approach we obtain the proper relationships
between the original BM submodels, especially that between the
$\gamma$-unstable Wilets-Jean and rigid rotor models. This is in
contrast to other phenomenological and microscopic approaches aiming
the microscopic foundation of the BM model. This fact was stressed
in Ref.\cite{RoweThiamova05} and was a central point in our
considerations. In some respects, the present considerations also
resemble those performed in the ACM. However, the main difference
between the present approach and the ACM is in the irreducible
collective subspaces in which the model Hamiltonians act. For the
microscopic models, like the PNSM, the state space is defined by
allowed $O(A-1)$ (or complementary to it $Sp(12,R)$) irreducible
representations $\omega$ that are consistent with the Pauli
principle, whereas for the phenomenological models the state space
in which the collective Hamiltonians act is defined by the
$O(A-1)$-scalar subspace of the many-particle Hilbert spaces with
$\omega = (0)$. The specific structure of this violated
permutational symmetry space $\mathbb{H}^{\omega=(0)}$ is that it
gives a "deep freezing" of the microscopic collective features of
the used Hamiltonians and make them similar to those in the
Bohr-Mottelson theory, associated with the irrotational-flow
collective dynamics. Thus, the combined proton-neutron dynamics in
the present approach is governed by the microscopic shell-model
intrinsic structure of the symplectic bandhead. To illustrate the
present symplectic-based proton-neutron shell-model approach, we
apply the theory to three nuclei with different collective
properties $-$ namely, $^{158}$Gd, $^{150}$Nd, and $^{148}$Nd. A
good description of the excitation energies of the ground, $\beta$
and $\gamma$ bands, as well as for the ground state intraband
$B(E2)$ transition strengths is obtained for these three nuclei. The
quadrupole collectivity is described without the use of an effective
charge. More detailed calculations will be given elsewhere.

\end{document}